\documentclass[a4paper,10pt]{article}
\usepackage{graphicx}
\usepackage{amssymb}
\usepackage{amsmath}
\usepackage{nicefrac}
\usepackage{dcolumn}
\usepackage{multirow}
\usepackage{cite}
\usepackage{mathrsfs}
\usepackage{bm}
\usepackage[usenames,dvipsnames]{xcolor}
\usepackage[colorlinks=true,citecolor=Blue,linkcolor=RubineRed,urlcolor=Blue]{hyperref}
\usepackage{bbm}
\usepackage{hyperref}
\usepackage{comment}
\usepackage{amsfonts}
\usepackage{float}
\usepackage{caption}
\usepackage{subcaption}
\usepackage{graphicx}
\usepackage{dcolumn}
\usepackage{bm}
\usepackage{xcolor}
\usepackage{ulem}
\usepackage{comment,caption,cite}

\usepackage{xcolor}
\usepackage{soul}
\setstcolor{red}

\begin{document}

\huge

\begin{center}
Ionization by electron impacts and ionization potential depression
\end{center}

\vspace{0.5cm}

\large

\begin{center}
Djamel Benredjem$^{a,}$\footnote{djamel.benredjem@universite-paris-saclay.fr}, Jean-Christophe Pain$^{b,c}$, Annette Calisti$^{d}$ and Sandrine Ferri$^{d}$
\end{center}

\normalsize

\begin{center}
\it $^a$Universit\'e Paris-Saclay, CNRS, Laboratoire Aim\'e Cotton, F-91405 Orsay, France\\
\it $^b$CEA, DAM, DIF, F-91297 Arpajon, France\\
\it $^c$Universit\'e Paris-Saclay, CEA, Laboratoire Mati\`ere sous Conditions Extr\^emes,\\
\it 91680 Bruy\`eres-le-Ch\^atel, France\\
\it $^d$Aix-Marseille Université, CNRS, Physique des Interactions Ioniques et Moléculaires, UMR7345, Campus Saint Jerome, 13397 Marseille Cedex 20, France
\end{center}

\vspace{0.5cm}

\begin{abstract}
We calculate the cross-section of ionization by free-electron impacts in high or moderate density plasmas. We show that the so-called ionization potential depression (IPD) strongly affects the magnitude of the cross-section in the high-density domain. We use the well-known IPD formulas of Stewart-Pyatt and Ecker-Kr\"oll. A more recent approach based on classical molecular dynamics simulation is also investigated. The latter provides an alternative way to calculate IPD values. At near-solid densities the effects of the free-electron degeneracy should be investigated. The rates are then calculated within the Fermi-Dirac statistics.

We first use the semi-empirical formula of Lotz for ionization cross-section. The results may differ significantly from measured cross-sections or calculations with reliable atomic codes. Then, in a second step, we propose a new formula that combines the Lotz formula and a polynomial expansion in terms of the ratio of the energy of the incident electron and the ionization energy. The coefficients of the polynomial expansion are adjusted to fit the cross-section provided by robust atomic codes. A great advantage of the new formula is that it allows a fully analytical calculation of the ionization rate.

Our results are compared to experiments measuring IPDs, cross-sections and rate coefficients on aluminum at high and moderate densities and on Be-like CNO ions.
\end{abstract}

\section{Introduction}
The radiative properties of hot and dense plasmas are well described in collisional-radiative calculations only if the rates of the involved processes are reliable and rapidly estimated in order to allow extensive calculations. Among these processes, the electron-impact ionization (EII) is investigated in this work because it plays an important role at high densities.

Different analytical formulas of the EII cross-section have been used in the past. Among them, the semi-empirical formulas of Drawin \cite{Drawin1966}, Lotz \cite{Lotz1967a,Lotz1967b,Lotz1968} and Younger \cite{Younger1982} have been widely used. More recently  Bernshtam \textit{et al.} \cite{Bernshtam2000} proposed an empirical formula for the direct ionization cross-section, which is similar to the formula of Lotz. It involves two parameters that depend on the orbital quantum number of the initial state. The two parameters are adjusted to fit experimental results. Calculations of the total cross-section corresponding to direct EII channels of argon and iron show that the Bernshtam \textit{et al.} empirical formula is more satisfactory than the formula of Lotz. Other authors utilize different empirical formulas. Let us mention the additional work of Lotz \cite{Lotz1970} which involves three parameters that are determined from experimental data. Rudge and Schwartz \cite{Rudge1966} also use a formula with three parameters which could be evaluated by fitting experimental data or numerical results. Llovet \textit{et al.} \cite{Llovet2014} described the essentials of classical, semi-classical and quantum models, and made an extensive comparison of measured K, L, and M shells of all elements from hydrogen to einsteinium (see also the references therein).

As they depend on the ionization energy, the cross-section as well as the rate are strongly affected by the so-called ionization potential depression (IPD), in the high-density domain. To estimate the IPD, two models have been developed five decades ago by Stewart and Pyatt \cite{Stewart-Pyatt} and Ecker and Kr\"oll \cite{Ecker-Kroll}. In experiments performed at LCLS (Stanford) on aluminum \cite{Ciricosta2016,Ciricosta2012} the observation of the K-$\alpha$ fluorescence and the measurement of the position of the K-edge of ions show that the formula of Ecker and Kr\"{o}ll is more adequate than the formula of Stewart and Pyatt. Nevertheless, the agreement is not satisfactory for the highest ion charges, \textit{i.e.}, from O-like to Be-like aluminum (see Fig. 4 in Ref. \cite{Ciricosta2016}). On the other hand, in an experiment performed at the Orion laser system (UK) \cite{Hoarty2013}, with a plasma at higher temperatures [500-700] eV and densities in the range [1-10] g/cm$^3$, the aluminum K-shell spectrum shows a better agreement with calculations if one uses the Stewart-Pyatt IPD rather than the Ecker-Kr\"{o}ll one. These two main experiments have stimulated many theoretical investigations of IPD (see for instance Refs. \cite{Son2014,Vinko2014,Lin2017,Lin2019}, in particular using Density Functional Theory.

Recent calculations on continuum lowering in plasmas under solar-interior conditions \cite{Zeng2020} showed that the silicon IPD presents a good agreement with the measurements of Ciricosta \textit{et al.} \cite{Ciricosta2016} for low ion charges, $z=4-6$, but disagrees for $z=7-10$. 

A model based on classical molecular dynamics (CMD) was developed at Aix-Marseille University \cite{Calisti2009}. It is designed to deal with neutral mixtures composed of ions of the same atom with different charge states and electrons. Thanks to the choice of the soft ion-electron potential, it has been possible to implement an ionization/recombination protocol to control the plasma ion charge distribution and the trapping of electrons in the ion wells. The ionization/recombination process allows an instantaneous knowledge of the potential energy of the valence electron of an ion with a given charge which takes into account the effects of the whole surrounding plasma. A statistical average of these data leads to a straightforward definition of the IPD. At the density and temperature of the experiments at LCLS, the IPD obtained within this approach shows a better agreement with the Ecker-Kr\"{o}ll IPD than with the Stewart-Pyatt IPD \cite{Calisti2017}.

In Section \ref{Sec:IPD}, we calculate the IPD in an aluminum plasma at a mass density of 2.7 g/cm$^3$ and an electron thermal energy of 50 eV, and compare the approaches of Stewart-Pyatt, Ecker-Kr\"{o}ll and the one based on molecular dynamics. 

In Section \ref{Sec:XSection}, we calculate the EII cross-section, restricting ourselves to direct transitions. Other mechanisms such as those involving a collisional excitation followed by  an auto-ionization or by a collisional ionization will not be considered in this work. We use the semi-empirical formula proposed by Lotz \cite{Lotz1967a}. Our calculated cross-sections involve the IPD.

In Section \ref{Sec:Rate}, we investigate the degeneracy effect of the free electrons on the EII by calculating the rate coefficient within the Fermi-Dirac statistics and comparing the results to the coefficient obtained within the Maxwell-Boltzmann statistics. 

Because the Lotz formula is not always satisfactory, we propose in Sec. \ref{Sec:Fitting} a new cross-section expressed as a product of the Lotz formula and a polynomial expansion in terms of the ratio of the free-electron kinetic energy to the ionization energy. The coefficients of the polynomial expansion are adjusted so that the new cross-section fits the results of two atomic codes, namely FAC \cite{Gu2008} and HULLAC \cite{BarShalom2001}.

\section{Ionization potential depression}\label{Sec:IPD}
The two mostly used approaches of the ionization potential lowering are briefly presented. In the approach of Stewart and Pyatt the IPD of an ion of net charge $ze$ is expressed as an expansion with respect to the ratio of the Debye length $\lambda_D$ and ion radius $R_0$, \textit{i.e.}

\begin{equation}
    I_{\rm SP}(z)=\frac{3(z+1)e^2}{2R_0}\left\lbrace\left [ 1+\left (\frac{\lambda_D}{R_0}\right )^3\right ]^{2/3}-\left (\frac{\lambda_D}{R_0} \right )^2\right\rbrace ,
\end{equation}
where 
\begin{equation*}\lambda_D=\sqrt{\frac{kT}{4\pi(N_e+\sum_z N_{z}z^2)e^2}},\end{equation*} with $N_e$ and $N_{z}$ representing the electron density and the density of ions of charge $ze$, respectively. The ion radius is given by 
$R_0=3/(4\pi N_i)^{1/3}$ where $N_i$ is the ion density: $N_i=\sum_z N_{z}$.

We define $\overline{Z^q}$ for a positive integer $q$: 
\begin{equation*}
\overline{Z^q}=\sum_zp_zz^q,
\end{equation*}
where $p_z$ is the fraction of ions of charge $ze$. The value $q=1$ defines the average ion charge. The rhs member may be written $\displaystyle\sum_zN_zz^q/N_i $ giving
\begin{equation*}\sum_zN_zz^q=N_i\overline{Z^q}.\end{equation*}
When $q=2$ the above relation gives the contribution of ions to the Debye length.

The high-density limit of the Stewart-Pyatt formula:
\begin{equation}
    I_{\rm SP-HD}(z)=\frac{3(z+1)e^2}{2R_0}\label{SP-HD},
\end{equation}
is widely used in the literature.

The Ecker-Kr\"oll model provides the following IPD:
\begin{equation}
    I_{\rm EK}(z)=\frac{(z+1)e^2}{R_0}\left\{\begin{array}{ll}
    R_0/\lambda_D & \;\;\;\;\mathrm{if} \;\;\;\; N_{\rm cr}\ge N_i(1+\overline{Z})\\
    C(1+\overline{Z})^{1/3} & \;\;\;\;\mathrm{if} \;\;\;\; N_{\rm cr}< N_i(1+\overline{Z}),
    \end{array}\right.\label{Eq:IEK}
\end{equation}
where \begin{equation*} N_{\rm cr}=\frac{3}{4\pi}\left (\frac{kT}{Z^2e^2}\right )^3\end{equation*} is the critical density, with $Z$ the atomic number. The constant $C$ is determined by imposing the continuity of the IPD at the critical density, giving 
\begin{equation*}C=\left (\frac{R_0}{(1+\overline{Z})^{1/3}\lambda_D}\right )_{N_{\rm cr}}.\end{equation*}
 
 While $I_{\rm SP-HD}$ depends only on the density, $I_{\rm EK}$ also depends on the temperature through the average ion charge, but the variation is not important in our study.
 
The IPD measured at LCLS \cite{Ciricosta2012} showed a better agreement with the formula of Ecker and Kr\"oll provided $C=1$, than with the formula of Stewart and Pyatt. On the other hand, the experiment performed at the Orion laser showed that the Stewart and Pyatt IPD is more satisfactory when used in the FLYCHK code \cite{Chung2005} to predict the X-ray emission of an aluminum plasma at mass densities reaching 10 g/cm$^3$.
 
\begin{figure}[ht!]
\begin{center}
\includegraphics[scale=0.6]{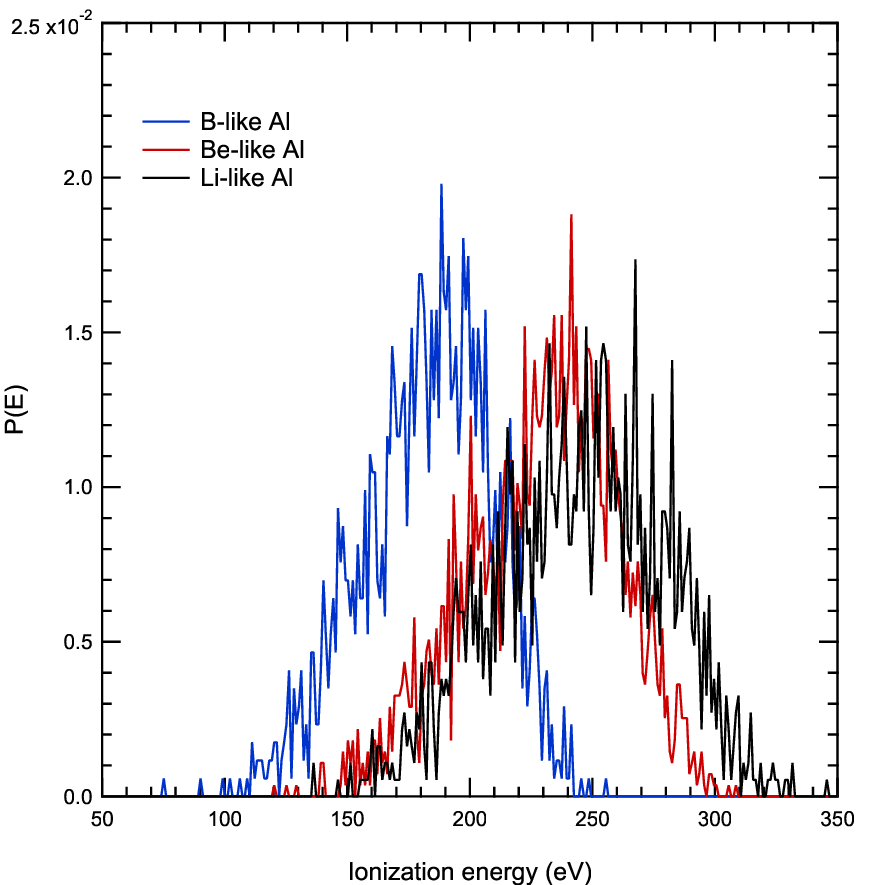}
\end{center}
\caption{Normalized CMD distribution of the ionization energy for Li-, Be- and B-like aluminum. Density=2.7 g/cm$^3$, $kT_e=50$ eV and $kT_i=300$ K.}
\label{Proba}
\end{figure}

The recent CMD approach consists in the simulation of the movement of interacting atoms or molecules treated as classical non relativistic point-like particles (for more details see for example Ref. \cite{Rapaport} and references therein). The Bingo-TCP code, used in the present study, has been designed to deal with neutral mixtures composed of ions of various charges and electrons and to allow the ion charges to change from one to another according to the density-temperature conditions. For that purpose, a regularized electron-ion potential, depending on the ion charge $ze$ is defined as:
\begin{equation}
V_{ie}(r)=-ze^2 e^{-r/\lambda}(1-e^{-r/\delta(z)})/r,
\end{equation}
where the regularization distance $\delta (z)$ is chosen to reproduce the ionization energy $E_z$ of the unperturbed ion of charge $ze$ in the ground state when the electron is located at the ion ($r=0$). Note that $\delta(z)$ is also used to define an exclusion sphere around ions and referred to ion stage radius. 
\begin{equation}
\delta(z)=z e^2/E_z.
\end{equation}
The screening factor, $e^{-r/\lambda}$ where $\lambda$ is half the simulation box size, helps to smooth the small fluctuations of forces arising with the periodic boundary conditions.  It has been checked, here, that the results do not depend on the choice of $\lambda$ provided that the box size is large enough (a few times the natural plasma screening length). The choice of this regularized ion-electron potential allows the implementation of an ionization/recombination protocol to control the plasma ion charge distribution and the trapping of electrons in the ion wells. The main idea of the model is to extract from the simulated particle positions and velocities, a local characterization of the plasma around an ion “A” in order to infer if the conditions are favorable to a ionization or recombination of this ion. For that purpose, the mutual nearest neighbor, ${\rm NN_A}$ and the next nearest neighbor, ${\rm NNN_A}$, electrons of “A” are identified and traced. Their total energy is calculated accounting for the whole complexity of the potential energy surface around “A” including the ionization energy lowering at a local level due to the surrounding charges. A shell noted $S_A$, formed with the ${\rm NN_A}$ and ${\rm NNN_A}$ is defined as the nearest environment of “A” if ${\rm NN_A}$ is localized at a distance $d_A$ of “A” such as $\delta(z_A) < d_A < \sqrt{2}\,\delta(z_A)$. Depending on the total energy of the two neighbor electrons, the shell is labeled hot (positive energy favorable to ionization) or cold (negative energy favorable to recombination). A hot or cold shell around an ion results, respectively, into a pre-ionization, \textit{i.e.}, an increase by 1 of the ion charge and the appearance of one electron localized at the ion, or a recombination, \textit{i.e.}, a decrease by 1 of the ion charge and the removal of the nearest-neighbor electron with a transfer of the kinetic energy difference to the NNN. This local discontinuity over one time step is then accounted for by the whole system through a normal evolution. The pre-ionized state, \textit{i.e.}, an ion with a trapped electron can then be converted into an ionized state through multiple collisions. In this approach the ionization will be considered as completed when a new hot shell surrounds the ion opening the way to a further pre-ionization. In the mean-time the ion is considered as excited or multi-excited if there are more than one trapped electron in the ion potential. It is important to note that the coupling of electrons with radiation is ignored in our model and that the notion of discrete energy for the ionic excited states is replaced here by its continuous equivalent. During the initial step of equilibration, the system is driven toward equilibrium using a thermostat and is not supposed to be used for any measurements. Once the system has reached an equilibrium state, the happening of ionization/recombination process becomes far less frequent than it was in the equilibration step, giving one the ability to go further in the description of the charge interactions in plasma, accounting for mixtures of ions undergoing changes of their charge states.
In particular, taking advantage of the particular design of the ionization protocol, it is possible, when the ion is in a pre-ionization state, to measure the necessary energy to free an electron in the ground state of a given ion and this, by accounting for all the interactions with the surrounding plasma. Due to the fluctuating local environment of the ions, the ionization energy is then characterized by a distribution function. If one compares the mean energies deduced from these distribution functions (see Fig. \ref{Proba}) with the corresponding energies for the isolated ions, it is possible to infer the IPD due to the interactions with the environment. Iglesias and Sterne \cite{Iglesias2013} investigated the fluctuations of the number of free electrons -and consequently the ion sphere radius- and proposed simple analytical IPDs within the models of Stewart-Pyatt and Ecker-Kr\"oll.

We investigate an aluminum plasma at a mass density 2.7 g/cm$^3$ and an electronic temperature $T_e=50$ eV. A calculation of the average ion charge with FLYCHK code \cite{Chung2005} gives $\overline{Z}=5.77$ and $\overline{Z^2}=34.1$. We have $\lambda_D/R_0=0.215$ which means that the high density limit of the Stewart-Pyatt IPD is a good approximation. Following experimental considerations \cite{Ciricosta2012} we assume $C=1$ and use the modified Ecker-Kr{\"o}ll IPD, 
\begin{equation}
    I_{\rm mEK}=\frac{2}{3}(1+\overline{Z})^{1/3}I_{\rm SP-HD}.\label{EKm}
\end{equation}

\begin{figure}[h]
\begin{center}
\includegraphics[scale=.6]{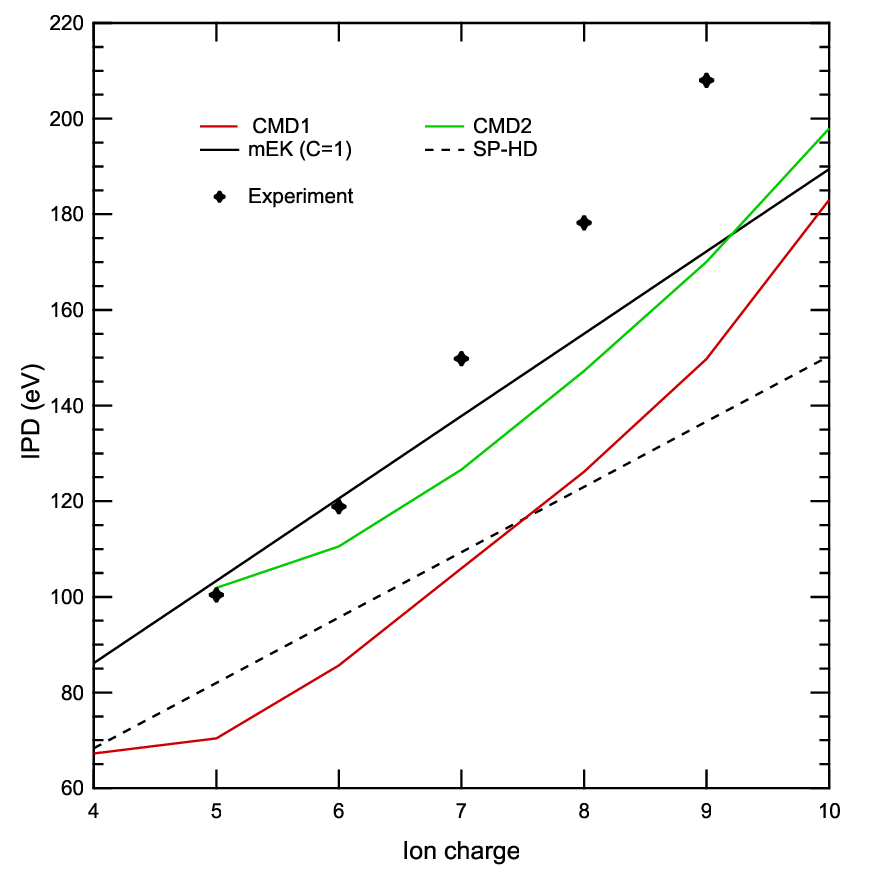}
\caption{Ionization potential depression in aluminum ions. CMD1 and CMD2: classical molecular dynamics with $T_i=50$ eV and 300 K, respectively ; mEK (C=1): modified Ecker-Kr{\"o}ll formula, with $C=1$ ; SP-HD: high density limit of the Stewart-Pyatt formula. Density=2.7 g/cm$^3$, $kT_e=50$ eV.}\label{IPD50}
\end{center}
\end{figure}

Figure \ref{IPD50} shows the calculated IPDs of aluminum ions having $z\in[4-10]$. The experimental IPD \cite{Ciricosta2012} is also represented. The CMD calculation with an ion temperature $T_i=300$ K (CMD2) shows a satisfactory agreement with the Ecker-Kr{\"o}ll IPD. Calculations (CMD2 and EK) agree with experimental results for the lowest ion charges only. 

As said above, a recent calculation on silicon under solar-interior conditions \cite{Zeng2020} shows the same divergence from experimental IPD for increasing $z$. We can see that the Stewart-Pyatt model as well as CMD1 calculation ($T_i=T_e$) are not satisfactory.  

The choice of a room temperature for $T_i$ is explained by the experimental conditions. In fact, the electrons in the target are heated within 80 fs to temperature up to 180 eV depending on the photon energy of the irradiation. Moreover, the K-shell fluorescence, on which the interpretation of the experiment is based, only occurs while the target is irradiated. On this time scale, the ion motion is negligible and emission occurs in a plasma at solid density. To get closer to these conditions one can use CMD to simulate a two-component plasma of ions at room temperature and solid density, and electrons in pseudo equilibrium with the cold-ion population.

As the continuum lowering is important at this density, it will be taken into account in cross-section calculations.

\section{Ionization cross-section}\label{Sec:XSection}
We first use the empirical formula of the cross-section proposed by Lotz \cite{Lotz1967a}. Other authors used similar formulas (see for example the work of Bernshtam and co-workers \cite{Bernshtam2000}). The cross-section of the direct ionization between the ground levels $g$ and $g'$ of ions of charges $ze$ and $(z+1)e$, respectively, reads
\begin{equation}
    \sigma(E)=A\,\xi \frac{\ln(E/E_{z,g})}{E E_{z,g}},\label{XSection}
\end{equation}
where $E$ is the incident electron energy, $A$ a constant (in the range $2.9-4.5\times 10^{-14}$ $\displaystyle {\rm cm^2\cdot eV^2}$, see Ref. \cite{Lotz1967a}), $E_{z,g}$ the ionization energy and $\xi$ the number of electrons in the subshell from which the ionization occurs. The ionization energy accounts for the continuum lowering, resulting in an increase of the cross-section. 

In fact, owing to a fluctuating environment of the ions, the classical molecular dynamics approach provides a distribution of ionization energy. As a consequence, $E_{z,g}$ is taken to be the average ionization energy. Therefore, this enables one to determine, not only the average ionization energy, but also the standard deviation of the latter, and all of its moments. For instance, the third- (skewness) and fourth- (kurtosis) order moments provide respectively information on the asymmetry and sharpness of the distribution. This opens the way to a statistical modeling of the ionization-energy distribution. Moreover, using the Bingo ionization-energy distribution, it is possible to average the whole cross-section itself, which could yield different results, compared to the procedure used in the present work consisting in including the average ionization energy in the cross-section. 

\begin{figure}
\centering
\includegraphics[scale=0.6]{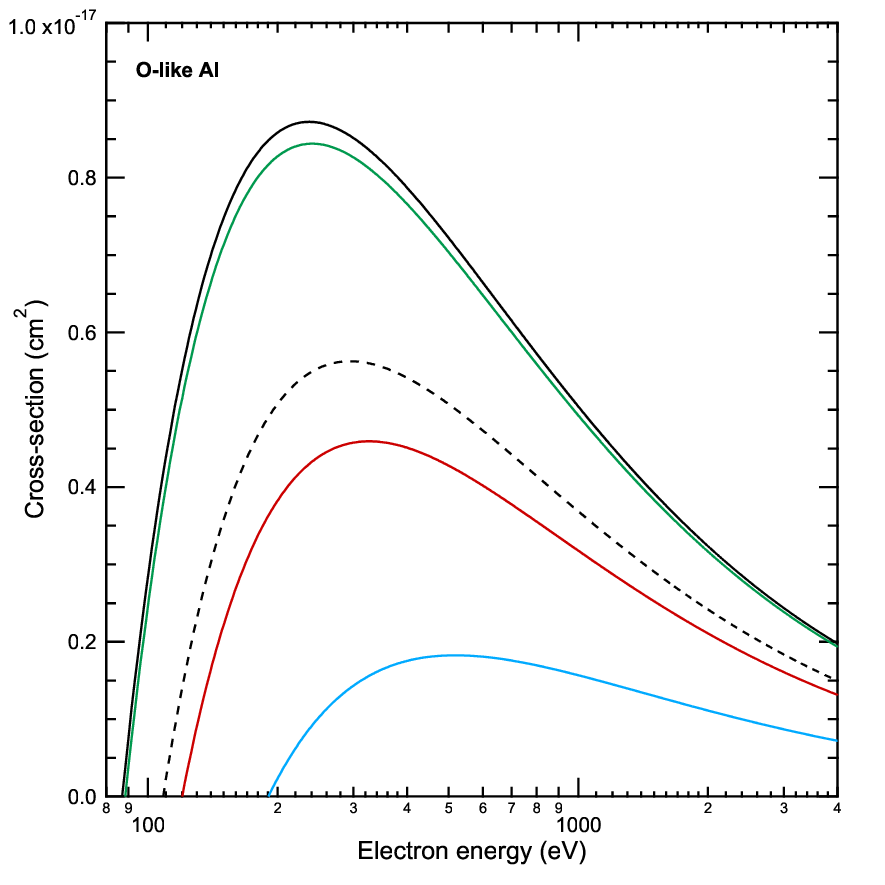}
  \caption{EII cross-section of O-like aluminum as a function of the energy of the incident electron. Legend, density and electron temperature: as in Fig. \ref{IPD50}. Blue curve: isolated ion. \label{XSection-O-like}}
\end{figure} 

\begin{figure} 
\centering
  \includegraphics[scale=0.6]{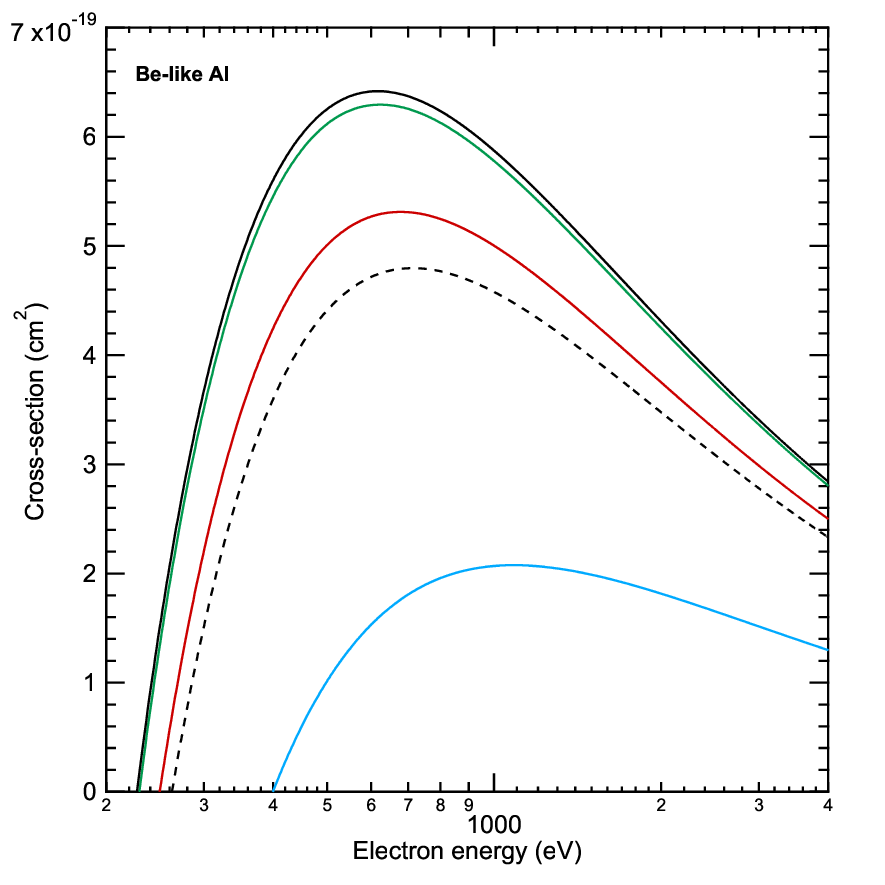}
  \caption{Same as in Fig. \ref{XSection-O-like} for Be-like aluminum. \label{XSection-Be-like}} 
\end{figure} 

In Fig. \ref{XSection-O-like} we show the EII cross-section of O-like aluminum. As can be seen from Fig. \ref{IPD50} the EK and CMD2 IPDs are very close to 100 eV, while the SP IPD is around 80 eV. Then, as expected, when we take EK or CMD IPDs the cross-sections are very close. By comparing with the isolated ion case it is clear that the IPD is responsible for a large increase of the cross-section. 

In Fig. \ref{XSection-Be-like} we represent the cross-section of Be-like aluminum. The values are smaller than those of the O-like ion by an order of magnitude. Moreover the difference between the various cross-sections is smaller than in the previous case. 

The Lotz formula allows us to derive an analytical form of the rate coefficients for both Fermi-Dirac statistics and Boltzmann distribution. The obtained rates are then more suitable when one deals with collisional-radiative equations in which case extensive calculations are required.
 
\section{Ionization rate}\label{Sec:Rate}
The ionization rate coefficient reads
\begin{equation}
    q=\int_{E_{z,g}}^{\infty}\sigma(E) \sqrt{\frac{2E}{m}}\,\rho(E)\,dE,\label{Rate}
\end{equation}
where $\sqrt{2E/m}$ is the electron velocity with $m$ the electron mass. The probability density $\rho(E)$ is expressed within the Fermi-Dirac statistics in order to account for the free-electron degeneracy.

\subsection{Fermi-Dirac statistics}
Let us introduce the Fermi-Dirac integral of integer and half-integer order $p$:
\begin{equation}
    F_{p}(\eta,\chi)=\frac{1}{\Gamma(p+1)}\int_{\chi}^{\infty}\frac{\epsilon^p}{e^{\epsilon-\eta}+1}d\epsilon,\label{FDintegral}
\end{equation}
where $\Gamma$ is the Gamma function. $\chi$ and $\eta$ are the reduced energy and chemical potential, respectively: $\epsilon=E/(kT)$ and $\eta=\mu/(kT)$. If $\chi\neq 0$, the integral is known as the incomplete Fermi-Dirac integral.

The normalized probability density is given by 
\begin{equation}
    \rho(E)=\frac{1}{D}\frac{\sqrt{E}}{e^{(E-\mu)/kT}+1},\label{FDD}  
\end{equation}
where $D=(kT)^{3/2}\Gamma(3/2)F_{1/2}(\eta,0)$. The second factor in the rhs is due to the normalization of $\rho$. We set $\chi=E_{z,g}/(kT)$. Equations (\ref{XSection}) and (\ref{Rate}) then become:
\begin{equation}
    \sigma=A\,\xi\frac{\ln(\epsilon/\chi)}{E\, E_{z,g}}\label{XSection2}
\end{equation}
and
\begin{equation}
    q=\sqrt{\frac{2}{m}}\frac{A\,\xi}{D}\frac{1}{\chi}\int_{\chi}^{\infty}\frac{\ln(\epsilon/\chi)}{e^{\epsilon-\eta}+1}d\epsilon.\label{Rate2}
\end{equation}

Let us focus on the integral in the equation above. We can write
$$
   \int_{\chi}^{\infty}\frac{\ln(\epsilon/\chi)}{e^{\epsilon-\eta}+1}d\epsilon=\ln\left (\frac{1}{\chi}\right )\Gamma(1)\,F_0(\eta,\chi)
   +\int_{\chi}^{\infty}\frac{\ln(\epsilon)}{e^{\epsilon-\eta}+1}d\epsilon,
$$
where $\Gamma(1)=1$. The incomplete Fermi-Dirac integral, $F_0$, can be expressed analytically \cite{Goano1995}, giving the first term in the rhs of the equation above:

\begin{equation*} 
I_1=\ln\left(\frac{1}{\chi}\right)F_0(\eta,\chi)=\ln\left(\frac{1}{\chi}\right)\lbrace\ln\left [e^{\chi-\eta}+1\right ]-(\chi-\eta)\rbrace.
\end{equation*}

The second integral
\begin{equation*}I_2=\int_{\chi}^{\infty}\frac{\ln(\epsilon)}{e^{\epsilon-\eta}+1}d\epsilon,\end{equation*}
is calculated numerically. Such an integral can also be estimated using a Sommerfeld-type expansion (see \ref{Appendix}). Thus the ionization rate coefficient becomes
\begin{equation}
    q=\sqrt{\frac{2}{m}}\frac{A\,\xi}{D}\frac{1}{\chi}(I_1+I_2)\label{Eq:q}
\end{equation}
or more explicitly 
\begin{equation}
    q=5.935\times 10^7\,A\,\xi\,\frac{1}{(kT)^{3/2}\,\Gamma(3/2)F_{1/2}(\eta,0)}\frac{1}{\chi}(I_1+I_2), \label{q}
\end{equation}
where the numerical factor is in $\displaystyle{\rm cm\cdot s^{-1}\cdot eV^{-1/2}}$ and $\Gamma(3/2)=\sqrt{\pi}/2$.

The chemical potential is obtained by
\begin{equation}
F_{1/2}(\eta,0)=\frac{(4\pi)^{3/2}}{2}\left (\frac{E_I}{kT}\right )^{3/2}N_e\,a_0^3,\label{chem.pot}
\end{equation}
where $a_0$ is the Bohr radius and $E_I$ the Rydberg energy. Knowing the electron density $N_e$ and temperature $T_e$ we calculate $F_{1/2}(\eta,0)$. It is then easy to derive the chemical potential. 

\begin{figure}[h] 
 \centering
  \includegraphics[scale=.6]{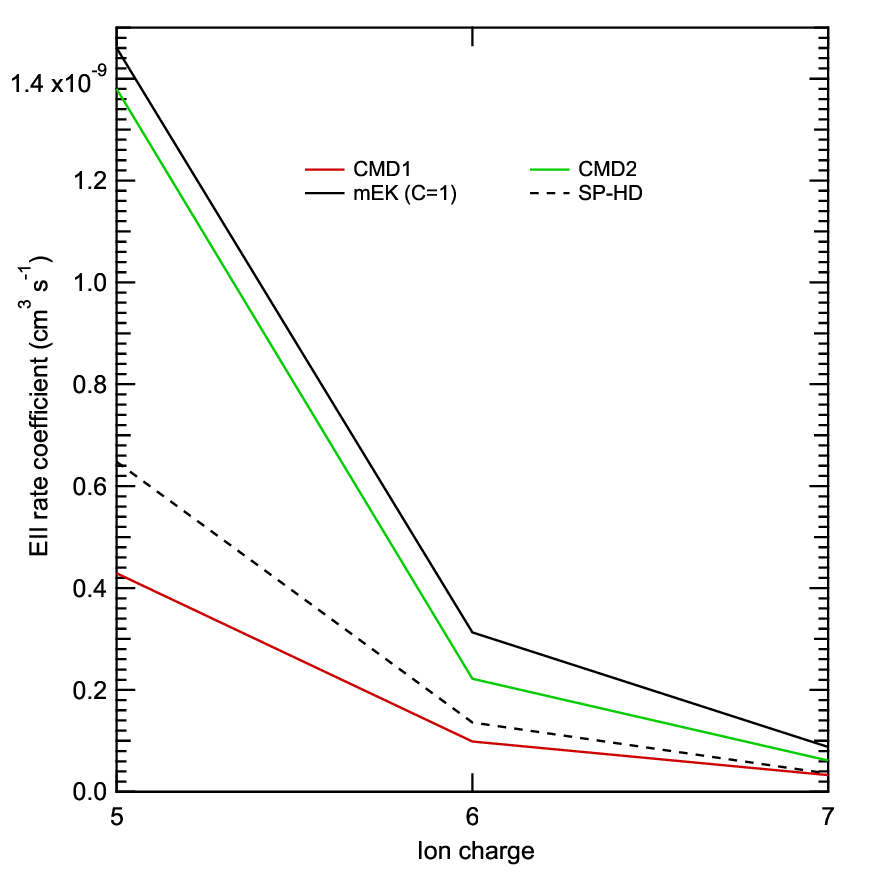}
  \caption{EII rate coefficient of aluminum ions, calculated within the Fermi-Dirac statistics for $z=5-7$. Legend, density and temperature, as in Fig. \ref{IPD50}. \label{Rate50eV_5to7}}
 \end{figure}
 
 \begin{figure}[h] 
 \centering
 \includegraphics[scale=0.6]{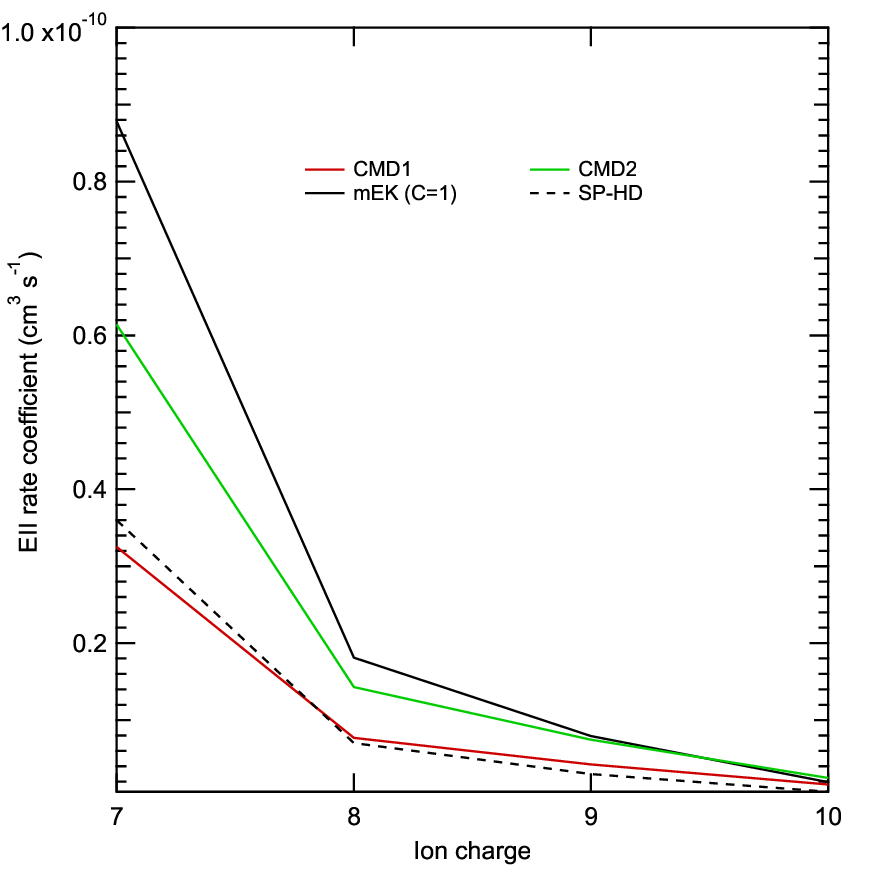}
  \caption{Same as in Fig. \ref{Rate50eV_5to7} for $z=7-10$. \label{Rate50eV_7to10}}
\end{figure} 

In Fig. \ref{Rate50eV_5to7} we show the EII rate coefficient of C-, N- and O-like aluminum. The rate coefficients calculated with the CMD2 IPD ($T_i=$300 K) show a good agreement with those obtained by using the EK IPD. When $T_i=T_e=50$ eV, the CMD1 IPD yields too low rates, even lower than the rates obtained with the high-density limit of the Stewart-Pyatt formula.

In Fig. \ref{Rate50eV_7to10} we represent the EII rate coefficient of C- to Li-like aluminum. We notice a larger discrepancy between CMD and EK results. However we have a good agreement between the rates using CMD1 and Stewart-Pyatt IPDs. 

\begin{figure}[h]
\centering
\includegraphics[scale=0.6]{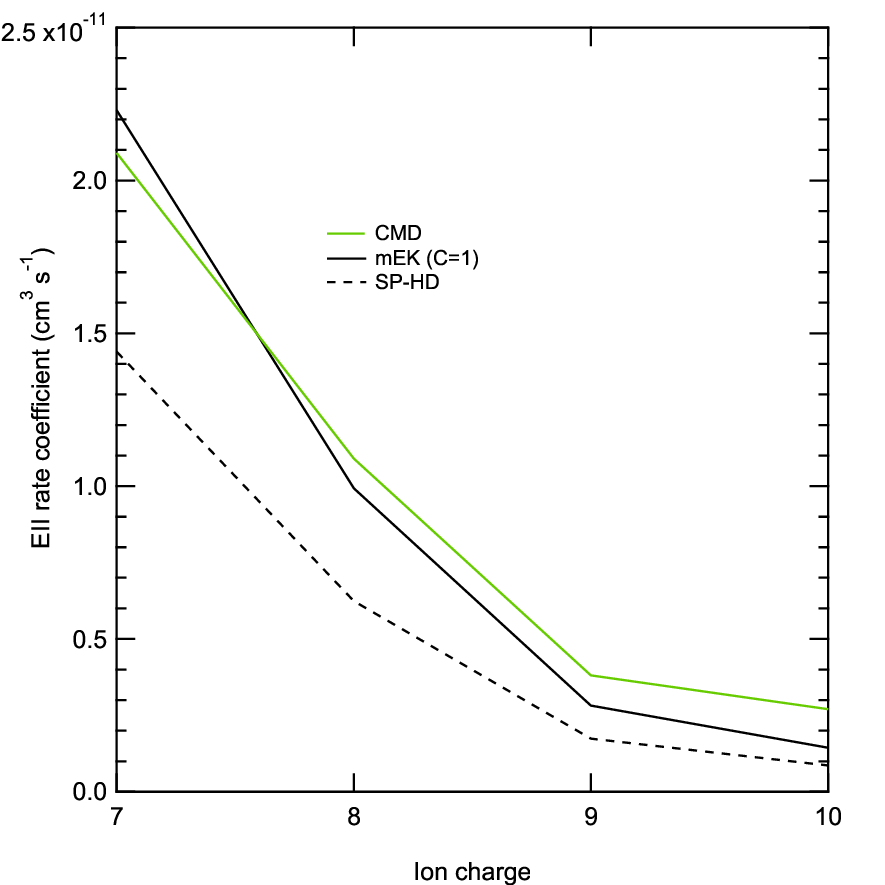}
 \caption{EII rate coefficient of aluminum ions.
$\rho$=0.34 g/cm$^3$, $T_e=T_i=70$ eV.\\Legend: as in Fig. \ref{IPD50}.\label{Rate034_70eV}}
 \end{figure}
 
 \begin{figure}[h]
 \centering
 \includegraphics[scale=0.6]{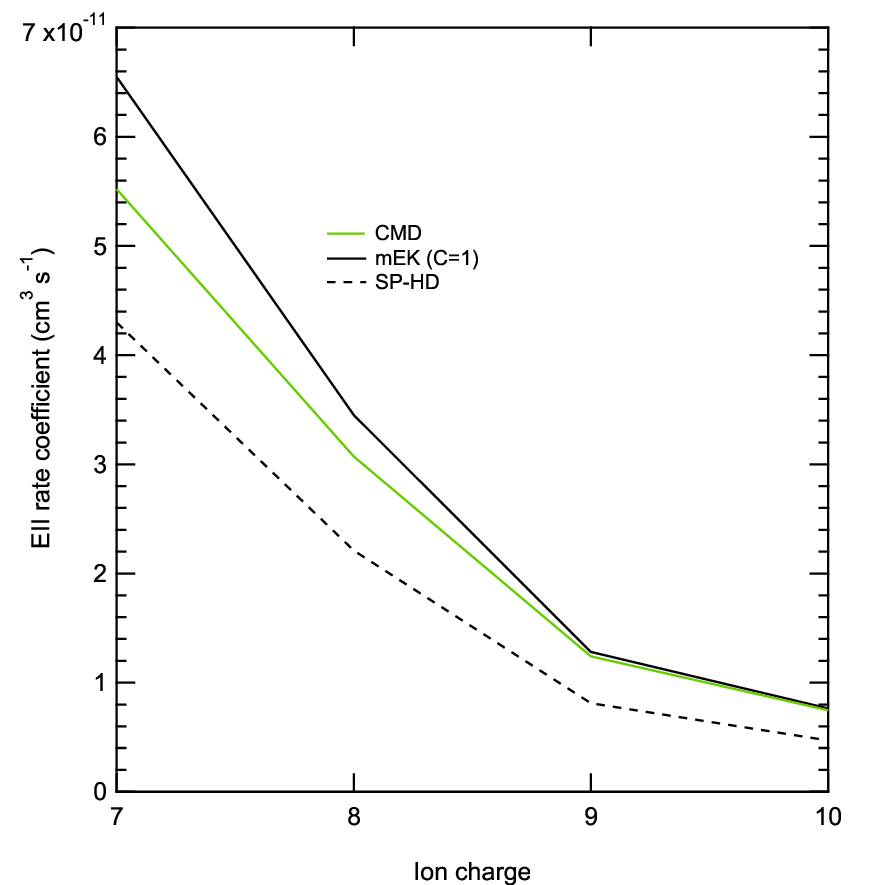}
\caption{Same as in Fig. \ref{Rate034_70eV}, with $T_e=T_i=100$ eV.\label{Rate034_100eV}}
 \end{figure} 

\begin{figure}[h]
\centering
\includegraphics[scale=0.6]{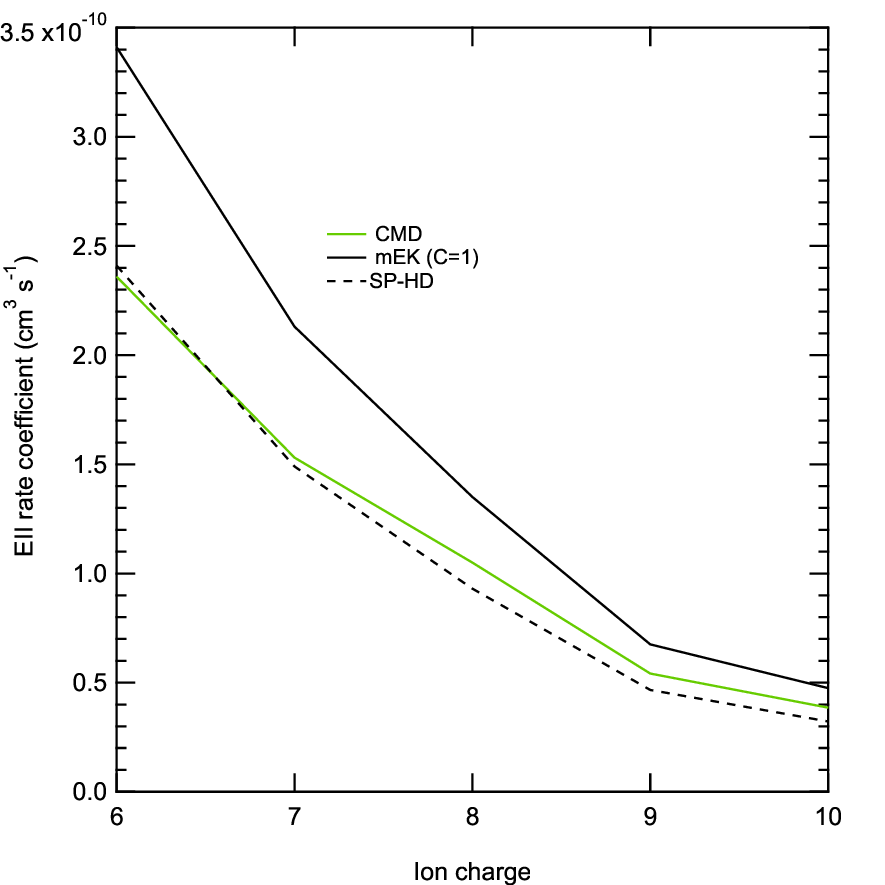}
\caption{Same as in Fig. \ref{Rate034_70eV}, with $T_e=T_i=190$ eV.}\label{Rate034_190eV}
\end{figure}

Now we show the variation of the rate coefficient with electron temperature. Here the mass density is fixed to  $0.34\ {\rm g/cm^3}$ and three electron temperatures are investigated. The ion temperature is identical to $T_e$. At 70 ev, the CMD and Ecker-Kr\"oll IPDs are close to each other. As a consequence, the rates calculated within the Fermi-Dirac statistics, and with these two IPDs, are very close (see (Fig. \ref{Rate034_70eV})). When the temperature is increased (see Figs. \ref{Rate034_100eV}-\ref{Rate034_190eV}), the rates calculated with CMD IPD tend to the rates obtained with the high-density limit of the Stewart-Pyatt IPD.

We know that the electron degeneracy effect is small when the electron temperature is well above the Fermi temperature. A direct consequence is that the Fermi-Dirac distribution tends towards the Boltzmann distribution. In the following, we show that the Lotz formula, coupled to a Boltzmann energy distribution of the free electrons, provides an analytical rate coefficient.

\subsection{Boltzmann energy distribution}
At the Boltzmann limit, the factor $\displaystyle \sqrt{E}/\left(e^{(E-\mu)/kT}+1\right )$ in Eq. (\ref{FDD}) becomes $\sqrt{E}\,e^{-(E-\mu)/kT}$. As a result, the product $\Gamma(3/2)F_{1/2}(\eta, 0)$ is easily calculated:
\begin{equation*}
\Gamma(3/2)F_{1/2}(\eta,0)\to e^{\eta}\int_0^{\infty}\epsilon^{1/2}e^{-\epsilon}d\epsilon
=\frac{\sqrt{\pi}}{2}e^{\eta}.
\end{equation*}

The Boltzmann distribution then reads
\begin{equation}
    \rho'(E)=\frac{\sqrt{E}\,e^{-E/kT}}{\frac{\sqrt{\pi}}{2}(kT)^{3/2}}.\label{MBD}
\end{equation}
The term $D$ in Eq. (\ref{FDD}) becomes 
\begin{equation}
    D'=\frac{\sqrt{\pi}}{2}(kT)^{3/2}e^{\eta}\label{Dprime}
\end{equation}
while $I_1$ and $I_2$ are replaced by
\begin{equation*}
     I_1'=\ln\left (\frac{1}{\chi}\right )e^{\eta}\int_{\chi}^{\infty}e^{-\epsilon}d\epsilon
     =\ln\left (\frac{1}{\chi}\right )e^{\eta-\chi}
\end{equation*}
and
\begin{eqnarray*}
     I_2'&=&e^{\eta}\int_{\chi}^{\infty}\ln(\epsilon)e^{-\epsilon}d\epsilon=e^{\eta}\left [e^{-\chi}\ln(\chi)+\int_{\chi}^{\infty}\frac{e^{-\epsilon}}{\epsilon}d\epsilon\right ]\nonumber\\
     &=&e^{\eta}\left [e^{-\chi}\ln(\chi)+E_1(\chi)\right ],\\
\end{eqnarray*}
where $E_1$ is the exponential integral \cite{Abramowitz}:
\begin{equation*}E_1(\chi)=\int_{\chi}^{\infty}\frac{e^{-\epsilon}}{\epsilon}d\epsilon.\end{equation*}
We then obtain 
\begin{equation*}I_1'+I_2'=e^{\eta}\,E_1(\chi)\end{equation*}
and the ionization rate coefficient becomes
\begin{equation*}q'=\sqrt{\frac{2}{m}}\frac{A\,\xi}{D'}\frac{1}{\chi}(I_1'+I_2'),\end{equation*}
and finally
\begin{equation}q'=\frac{4}{\sqrt{2\pi m}}\frac{A\,\xi}{(kT)^{3/2}}\frac{E_1(\chi)}{\chi}
    \label{qprime1}
\end{equation}
or
\begin{equation*}q'=6.697\times10^7\frac{A\,\xi}{(kT)^{3/2}}\frac{E_1(\chi)}{\chi},\end{equation*}
where the numerical constant is given in $\rm{cm\cdot s^{-1}\cdot eV^{-1/2}}$.

\subsection{Comparison}
We compared the rates calculated with Eq. (\ref{q}) (Fermi-Dirac statistics) and Eq. (\ref{qprime1}) (Maxwell-Boltzmann statistics). The IPD value is given by a molecular-dynamics calculation. The difference between the two rates is very small, less than 7 \%, showing that the degeneracy of the free electrons plays a small role. In fact, at the density $\rho=2.7$ g/cm$^3$ and thermal energy $kT_e=50$ eV, the Fermi energy $E_{\rm F}=20.745$ eV while the chemical potential $\mu=-81.45$ eV. We then have $e^{-\mu/kT}=5.1$. The Maxwell-Boltzmann limit is then relevant. Figures 1 and 2 of Ref. \cite{Williams2020} confirm that our plasma is in the classical regime and consequently that the free-electron energy distributions (Eqs. (\ref{FDD}) and (\ref{MBD})) are very close. Free-electron degeneracy therefore plays a very small role on the ionization at the above conditions.

On the other hand, a recent work \cite{Tallents2016} shows that the degeneracy plays an important role during the interaction of an EUV free-electron laser with solid creating a warm density plasma ($kT_e<10$ eV).

\section{Numerical calculations}\label{Sec:Fitting}
The Lotz formula allows an easy analytical calculation of the rates. Unfortunately, sometimes, the results are very different from measured cross-sections. The alternative to the Lotz formula is a numerical calculation with a reliable atomic code. In this section, we rely on the two codes, FAC and HULLAC, to obtain accurate cross-sections. The resulting cross-sections are then compared to experimental results.

FAC is an integrated software package capable of investigating the atomic structure as well as most processes occurring in plasmas. It provides the level energies and the rate of the following processes: radiative transitions, collisional excitation, electron-impact ionization, photoionization, autoionization, radiative recombination and dielectronic capture. In this work, the EII cross-sections are computed in the distorted wave (DW) approximation. Bound and free states are determined via a self-consistent field model, and a local term for exchange is added to the potential (Dirac-Fock-Slater approach). The code also includes a collisional radiative model to construct synthetic spectra for plasmas under different physical conditions.

HULLAC is also an integrated computer package for atomic processes in plasmas. Like FAC it enables one to calculate atomic structure, cross-sections and rates for collisional and radiative atomic processes. The code is based on relativistic quantum-mechanical calculations including configuration interaction. The collisional cross-sections are calculated within the DW approach. The parametric potential method is used for both bound and free orbitals. The factorization–interpolation method is applied to the derivation of collisional rates. The continuum orbitals are computed in the framework of the phase-amplitude approach. The NJGRAF graphical method is used in the calculation of the angular momentum part of the matrix elements. Physics and code descriptions can be found in Ref. \cite{BarShalom2001}.

In the following, the densities of the investigated plasmas are much lower than in the experiments at LCLS and Orion laser. The ionization potential depression is then negligible. As a consequence, the cross-section and the rate involve the ionization energy of isolated ions.

\subsection{EII cross-section in Be-like CNO}
We have calculated the cross-section of ionization from the ground level of Be-like ions forming Li-like ions, with FAC and HULLAC codes. Our calculations are compared to measurements utilizing the crossed electron and ion beams technique at Oak Ridge National Laboratory \cite{Fogle2008}.

\begin{figure}[h]
 \centering
 \includegraphics[scale=0.6]{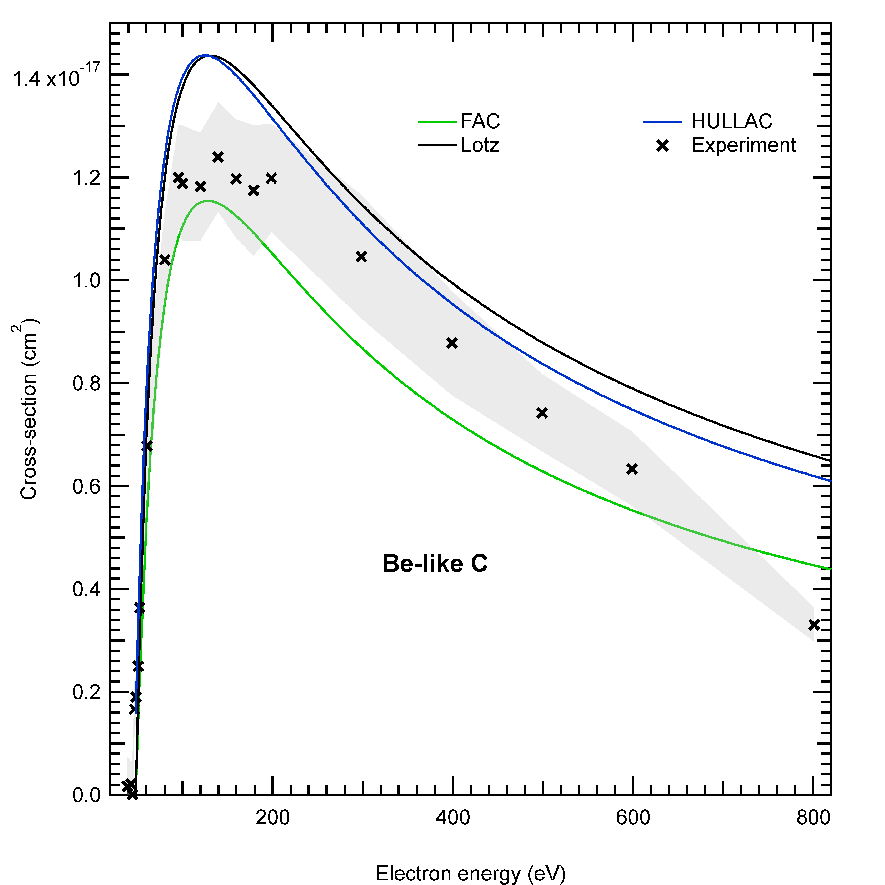}
  \caption{Cross-section of EII from the ground level of Be-like carbon. The gray surface represents the experimental error.}\label{Carbon}
 \end{figure}
 \begin{figure}[h]
 \centering
  \includegraphics[scale=0.6]{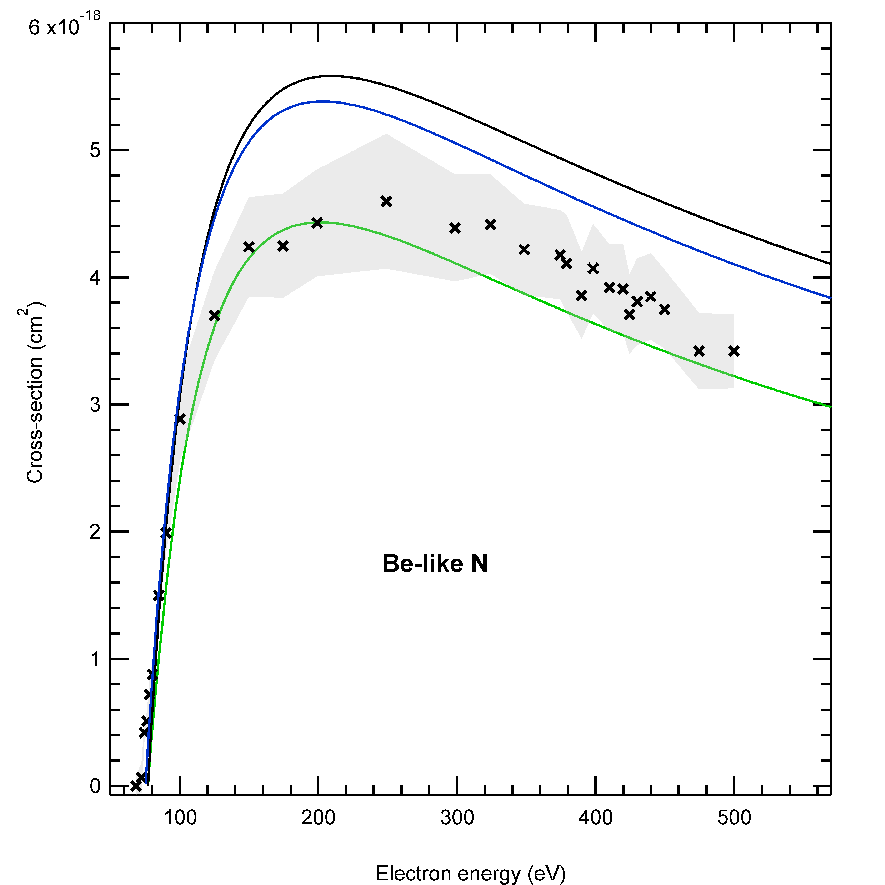}
  \caption{Same as in Fig. \ref{Carbon} for nitrogen.} \label{Nitrogen}
 \end{figure}

\begin{figure}[ht!]
\centering\includegraphics[scale=0.6]{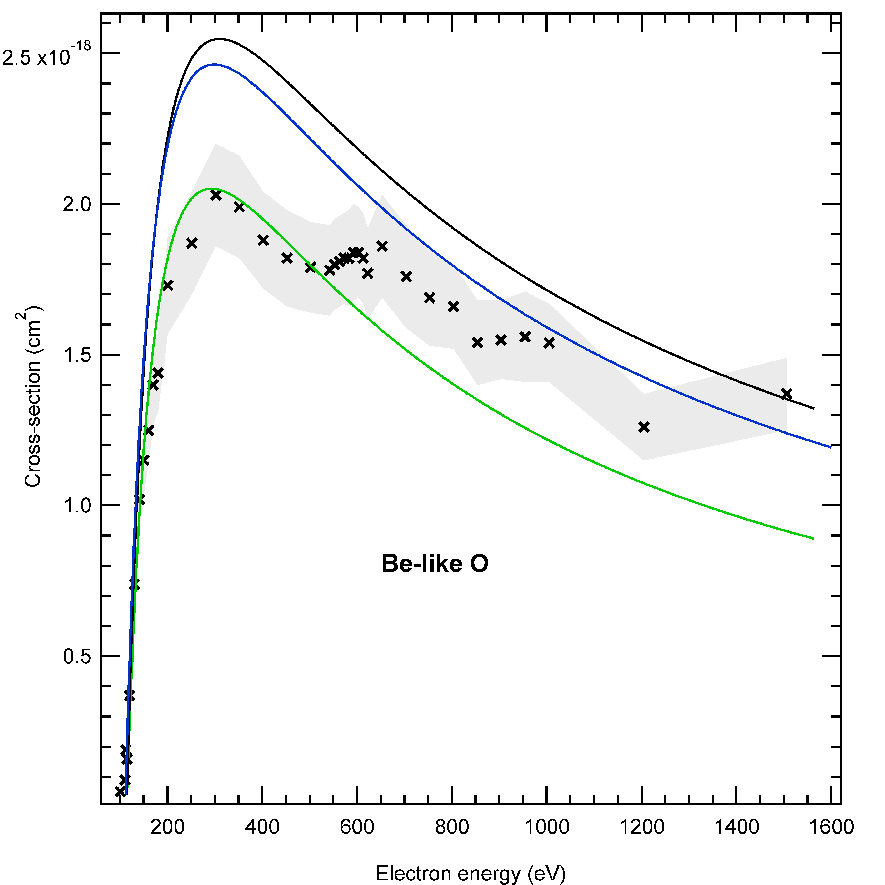}
\caption{\label{fig:Oxygen} Same as in Fig. \ref{Carbon} for oxygen.}\label{Oxygen}
\end{figure}

In Figs. \ref{Carbon}-\ref{Oxygen} we show the EII cross-section in Be-like carbon, nitrogen and oxygen, respectively. We consider only the ionization from the ground level to the Li-like levels $2s$ and $2p\ (J=1/2, 3/2)$. As clearly seen, the Lotz formula and the numerical (FAC and HULLAC) results have the same behaviour with respect to energy. However, the Lotz and HULLAC cross-sections differ by a significant amount from measurements for all ions, in almost the entire energy range. Our calculations with FAC show a better agreement with experimental results. This gives us confidence in the FAC code.

The ionization from the metastable states $2s2p$ ($J=0$, 1, 2) is not taken into account in this work. However, a preliminary calculation with FAC, and the fractions of the ground vs metastable states estimated by Fogle \textit{et al.} \cite{Fogle2008}, shows a small difference with ionization from the ground state only. The study of the metastable states is in progress and their contribution to the total cross-section will be addressed in a future work.

In the following, we present a new approach that gives the EII rate coefficient. As above the rate coefficient is calculated within the Fermi-Dirac statistics and at the Boltzmann limit. In contrast to Sec. \ref{Sec:Rate} we use a new approach utilizing both the accuracy of FAC code and an analytical calculation. The calculated rate is then compared to measurements.

\subsection{New rate coefficient. Comparison with experiment}
In this section we first compare the semi-empirical formula, Eq. (\ref{XSection2}), used by Lotz \cite{Lotz1967a} and Bernshtam \textit{et al.} \cite{Bernshtam2000} to cross-sections that are obtained with the FAC code \cite{Gu2008}. When the difference is significant we propose in a second step the following procedure: we replace the semi-empirical formula by a new formula that reproduces the FAC cross-sections. This procedure has the advantage to allow for an analytical calculation of the rate. To be specific the semi-empirical cross-section is multiplied by a polynomial expansion in which the coefficients are adjusted to fit with the FAC cross-section, \textit{i.e.} 
\begin{equation}
 \sigma= A\,\xi\frac{\ln(\epsilon/\chi)}{E\, E_{z,g}}\times \sum_{p=0}^Na_p\left (\frac{\epsilon}{\chi}\right )^p,\label{Expansion}
 \end{equation}
 where $N$ is the degree of the polynomial, $\epsilon/\chi=E/E_{z,g}$, \textit{i.e.} the ratio of the incident electron energy and the ionization energy. The fit of the FAC cross-section yields the $a_p$ coefficients. In the following, we show that the new rate coefficient can be expressed in terms of the $a_p$s.

\subsubsection{Fermi-Dirac distribution}
If the free electrons evolve according to the Fermi-Dirac distribution, Eq. (\ref{Rate2}) giving the rate reads
\begin{eqnarray}
q&=&\sqrt{\frac{2}{m}}\frac{A\,\xi}{D}\frac{1}{\chi}\int_{\chi}^{\infty}\frac{\ln(\epsilon/\chi)}{e^{\epsilon-\eta}+1}\sum_{p=0}^Na_p\left (\frac{\epsilon}{\chi}\right )^p\,d\epsilon=\nonumber\\
&=&\sqrt{\frac{2}{m}}\frac{A\,\xi}{D}\frac{1}{\chi}\sum_{p=0}^N\frac{a_p}{\chi^p} \int_{\chi}^{\infty}\frac{\ln(\epsilon/\chi)}{e^{\epsilon-\eta}+1}\epsilon^p\,d\epsilon\nonumber\\
&=&\sqrt{\frac{2}{m}}\frac{A\,\xi}{D}\frac{1}{\chi}\sum_{p=0}^N\frac{a_p}{\chi^p}\lbrack I_1^{(p)}+I_2^{(p)}\rbrack,
\label{qExpansion}
\end{eqnarray}
where $I_1^{(p)}$ can be expressed in terms of the incomplete Fermi-Dirac integral
\begin{equation*}
I_1^{(p)}=-\ln (\chi)\int_{\chi}^{\infty}\frac{\epsilon^p}{e^{\epsilon-\eta}+1}d\epsilon=-\ln(\chi)\Gamma(p)\,F_p(\eta,\chi)
\end{equation*}
and
\begin{equation*}I_2^{(p)}=\int_{\chi}^{\infty}\frac{\epsilon^p\,\ln \epsilon}{e^{\epsilon-\eta}+1}d\epsilon.\end{equation*}

\subsubsection{Boltzmann distribution}
In the case of the Maxwell-Boltzmann statistics, we have $I_1^{(p)}\to I_1'^{(p)}$ and $I_{2}^{(p)}\to I_2'^{(p)}$, with

\begin{equation*}
I_1'^{(p)}=-e^{\eta}\ln (\chi)\int_{\chi}^{\infty} \epsilon^p\,e^{-\epsilon}\,d\epsilon
=-e^{\eta}\ln (\chi)\Gamma(p+1,\chi),
\end{equation*}
where $\Gamma(p+1,\chi)$ is the incomplete Gamma function, and 
\begin{eqnarray*}
I_2'^{(p)}=e^{\eta}\int_{\chi}^{\infty}\epsilon^p\,e^{-\epsilon}\,\ln(\epsilon)\,d\epsilon.
\end{eqnarray*}
The $\Gamma$ functions can be calculated by using the relation $\Gamma(p+1,\chi)=p\Gamma(p,\chi)+\chi^p e^{-\chi}$, and by knowing that $\Gamma(1,\chi)=e^{-\chi}$.
It is then easy to obtain
\begin{eqnarray*}
I_1'^{(0)}&=&-e^{\eta-\chi}\ln(\chi)\\
I_1'^{(1)}&=&-e^{\eta-\chi}\ln(\chi)\lbrack\chi+1\rbrack\\
I_1'^{(2)}&=&-e^{\eta-\chi}\ln(\chi)\lbrack\chi^2+2\chi+2\rbrack\\
I_1'^{(3)}&=&-e^{\eta-\chi}\ln(\chi)\lbrack\chi^3+3\chi^2+6\chi+6\rbrack\\
\vdots
\end{eqnarray*}\\
The $I_2'^{(p)}$ are integrated by parts. We
have 

\begin{eqnarray*}
I_2'^{(p)}&=&e^{\eta}\left \lbrack \frac{\epsilon^{p+1}}{p+1}e^{-\epsilon}\ln(\epsilon)\right\rbrack^{\infty}_{\chi}-\frac{1}{p+1}\int_{\chi}^{\infty}\epsilon^{p+1}\left\lbrack-e^{-\epsilon}\ln(\epsilon)+\frac{e^{-\epsilon}}{\epsilon}\right\rbrack d\epsilon\nonumber\\
&=&\frac{e^{\eta}}{p+1}\left\lbrack-\chi^{p+1}e^{-\chi}\ln(\chi)+\int_{\chi}^{\infty}\epsilon^{p+1}e^{-\epsilon}\ln(\epsilon)d\epsilon-\int_{\chi}^{\infty}\epsilon^{p}e^{-\epsilon}d\epsilon\right\rbrack\nonumber\\
&=&\frac{1}{p+1}\left\lbrack-\chi^{p+1}e^{\eta-\chi}\ln(\chi)+I_2'^{(p+1)}-e^{\eta}\Gamma(p+1,\chi)\right\rbrack.
\end{eqnarray*}
It is then easy to write $I_2'^{(p+1)}$ in terms of $I_2'^{(p)}$:
\begin{equation}
    I_2'^{(p+1)}=(p+1)I_2'^{(p)}+e^{\eta}\Gamma(p+1,\chi)+\chi^{p+1}e^{\eta-\chi}\ln(\chi).
\end{equation}

We calculate the $p=0$ integral and deduce the higher-order ones. We have

\begin{eqnarray*}
I_2'^{(0)}&=&e^{\eta}\int_{\chi}^{\infty}e^{-\epsilon}\ln(\epsilon)d\epsilon
=e^{\eta}\left (\left\lbrack -e^{-\epsilon}\ln(\epsilon)\right\rbrack_{\chi}^{\infty}+\int_{\chi}^{\infty}\frac{e^{-\epsilon}}{\epsilon}d\epsilon\right )\\
&=&e^{\eta}E_1(\chi)+e^{\eta-\chi}\ln(\chi)
\end{eqnarray*}
and
\begin{eqnarray*}
I_2'^{(1)}&=&e^{\eta} E_1(\chi)+e^{\eta-\chi}[1+(1+\chi)\ln(\chi)]\\
I_2'^{(2)}&=&2e^{\eta}E_1(\chi)+e^{\eta-\chi}\lbrack \chi+3+(\chi^2+2\chi+2)\ln(\chi)\rbrack\\
I_2'^{(3)}&=&6e^{\eta}E_1(\chi)+e^{\eta-\chi}\lbrack \chi^2+5\chi+11
+(\chi^3+3\chi^2+6\chi+6)\ln(\chi)\rbrack.\\
\vdots
\end{eqnarray*}

The first $I_1'^{(p)}+I_2'^{(p)}$ sums are then 

\begin{eqnarray*}
I_1'^{(0)}+I_2'^{(0)}&=&e^{\eta}E_1(\chi)\\
I_1'^{(1)}+I_2'^{(1)}&=&e^{\eta}[E_1(\chi)+e^{-\chi}]\\
I_1'^{(2)}+I_2'^{(2)}&=&e^{\eta}[2E_1(\chi)+e^{-\chi}(3+\chi)]\\
I_1'^{(3)}+I_2'^{(3)}&=&e^{\eta}[6E_1(\chi)+e^{-\chi}(\chi^2+5\chi+11)]\\
\vdots
\end{eqnarray*}

The rate coefficient is then given by Eqs. (\ref{qExpansion}) replacing $I_1^{p}+I_2^{p}$ by $I_1'^{(p)}+I_2'^{(p)}$ and $D$ by $D'$ (see Eq. (\ref{Dprime})). We then have
\begin{equation}
    q'=6.697\times 10^7\frac{A\,\xi}{(kT)^{3/2}}\frac{e^{-\eta}}{\chi}\sum_{p=0}^N\frac{a_p}{\chi^p}[I_1'^{(p)}+I_2'^{(p)}].\label{qprimeExpansion}
\end{equation}

\subsubsection{Comparison with experiment}
In the following, we compare our calculations on aluminum ions to the measurements of Greve \textit{et al.} \cite{Greve1982}. In this experiment, aluminum and silicon ions were introduced in a well-diagnosed theta-pinch discharge by CO$_2$ laser driven ablation from solid targets. The authors interpreted the time histories of spectral lines from several ionization stages of these impurities, produced in the hot transient pinch plasma, in terms of effective ionization rate coefficients. 

The measured electron densities and temperatures are reported in Table \ref{Plasma conditions}. We also give the reduced chemical potential (Eq. (\ref{chem.pot})) and ionization energy as well as the the ratio of the SP and EK IPD values to the ionization energy. We can see that $\chi$ ranges in the interval $[1.5,2]$ and that $\eta\simeq -20$, showing that the Boltzmann distribution describes very well the free electrons. The average ion charge is given by FLYCHK code \cite{Chung2005}. We have $\overline{Z}\simeq 11$ for most ions. Due to the low density values, the IPD is negligible with respect to the ionization energy. As a result, the wavefunctions and cross-sections are not affected by plasma density effects.

\begin{center}
\begin{table}[h]
\caption{Plasma status in the experiment of Greve \cite{Greve1982}. $N_e$ in cm$^{-3}$, $kT_e$ in eV. $\eta$ and $\chi$ are the reduced chemical potential and ionization energy, \textit{i.e.} $\eta=\mu/(kT_e)$, $\chi=E_{z,g}/(kT_e)$. $\bar{Z}$ is the average ion charge. SP-IPD/$E_{z,g}$ (EK-IPD/$E_{z,g}$) is the ratio of the IPD calculated with the formula of Stewart-Pyatt (Ecker-Kr\"oll) and the ionization energy.}
\centering
\begin{tabular}{cccccccc}\hline\hline
& $kT_e$ & $N_e$ & $\eta$ &$\chi$&$\overline{Z}$ &SP-IPD/$E_{z,g}$ &EK-IPD/$E_{z,g}$ \\\hline\hline
Li-like & 225 & 3.2$\times 10^{16}$ & -20.15 & 1.97 & 10.9 &1.24 $\times 10^{-3}$&3.43 $\times 10^{-3}$\\\hline
Be-like & 235 & 2.7$\times 10^{16}$ & -20.39 & 1.69 & 10.9 &1.19 $\times 10^{-3}$ &3.83 $\times 10^{-3}$ \\\hline
B-like  & 220 & 2.1$\times 10^{16}$ & -20.54 & 1.5 & 10.9 & 1.18 $\times 10^{-3}$ &4.61 $\times 10^{-3}$ \\\hline
C-like  & 175 & 1.5$\times 10^{16}$ & -20.53 & 1.62 & 10.6 &1.10 $\times 10^{-3}$ &5.32 $\times 10^{-3}$ \\\hline
N-like  & 160 & 1.3$\times 10^{16}$ & -20.54 & 1.50 & 10.4 &1.09 $\times 10^{-3}$ &6.25 $\times 10^{-3}$ \\\hline\hline
\end{tabular}\label{Plasma conditions}
\end{table}
\end{center}

\begin{figure}[h]
\centering
  \includegraphics[scale=0.6]{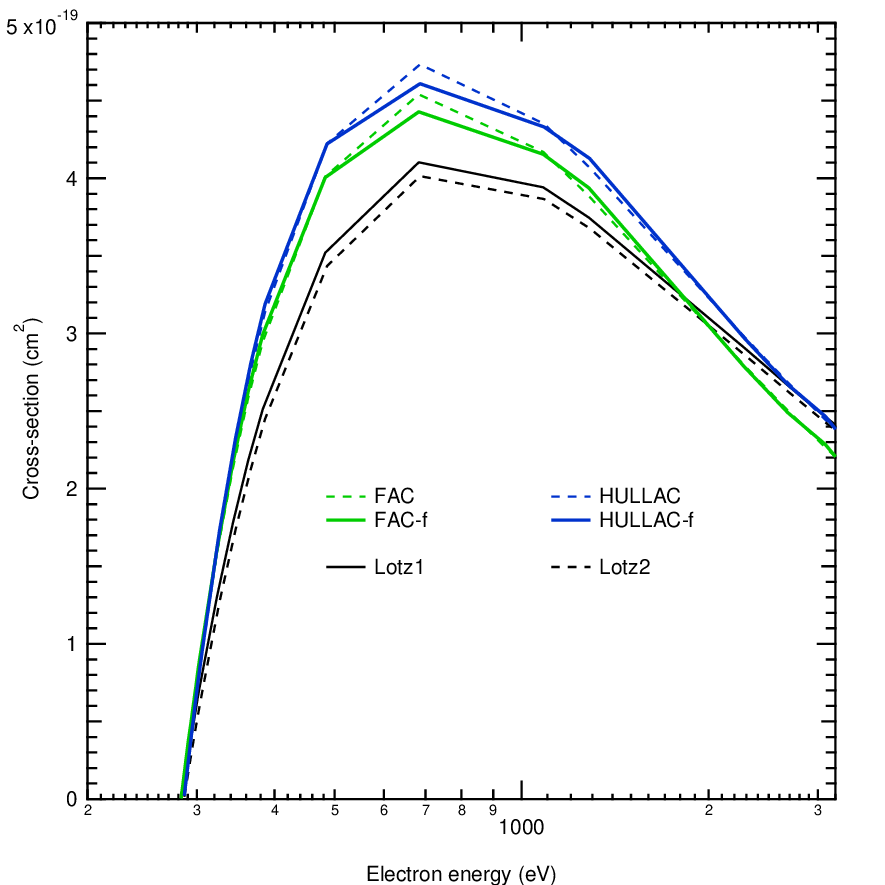}
  \caption{EII cross-section of C-like aluminum as a function of the energy of the incident electron. Density and temperature: 1.5$\times 10^{16}$ cm$^{-3}$ and 175 eV (see Table \ref{Plasma conditions}). FAC and HULLAC curves: numerical results given by FAC and HULLAC codes, respectively; FAC-f and HULLAC-f: our calculations, with the {$a_p$} coefficients obtained by a fit of the new cross-section (Eq. (\ref{Expansion})) with FAC or HULLAC cross-sections ; Lotz1 and Lotz2: Lotz cross-section with ionization energy given by FAC and HULLAC codes, respectively.\label{XSection_C-like_Al}}
 \end{figure}

The cross-sections are given by Eq. (\ref{Expansion}) where the $a_p$ coefficients are obtained by a fit with the FAC or HULLAC cross-sections. 
In Fig. \ref{XSection_C-like_Al} we show the cross-section of ionization from the ground state of C-like aluminum to the $2p$ states of B-like aluminum. FAC and HULLAC give similar cross-sections. Nevertheless, the two codes show a significant difference with the Lotz cross-section. Our fit of the FAC and HULLAC cross-sections is satisfactory. The obtained $a_p$ values are then expected to provide rates in agreement with experimental results. 

\begin{figure}[h]
 \centering
 \includegraphics[scale=0.6]{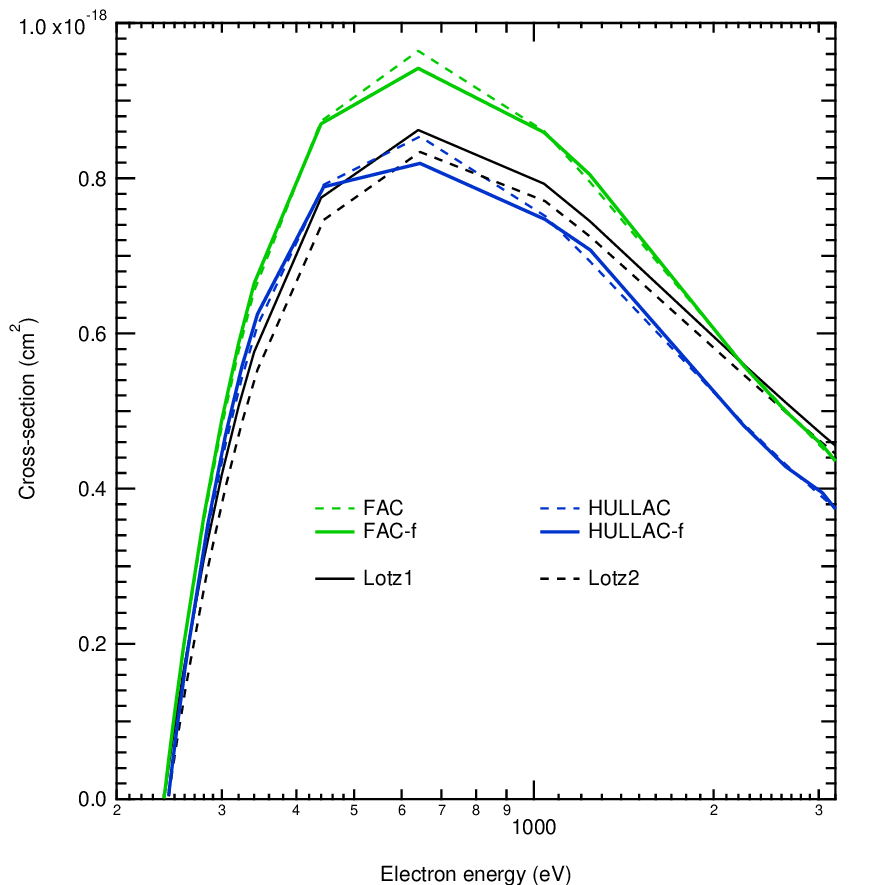} 
  \caption{EII cross-section of N-like aluminum as a function of the energy of the incident electron. Density and temperature: 1.3$\times 10^{16}$ cm$^{-3}$ and 160 eV (see Ref. \cite{Greve1982}). Legend, as in Fig. \ref{XSection_C-like_Al}. \label{XSection_N-like_Al}} 
\end{figure}

Similarly, Fig. \ref{XSection_N-like_Al} shows the cross-section of ionization from N-like aluminum. The difference between FAC and HULLAC is larger than in the C-like aluminum case. More interesting, the HULLAC results show a better agreement with the Lotz formula than with the FAC results.


\begin{figure}[t]
\centering\includegraphics[scale=0.6]{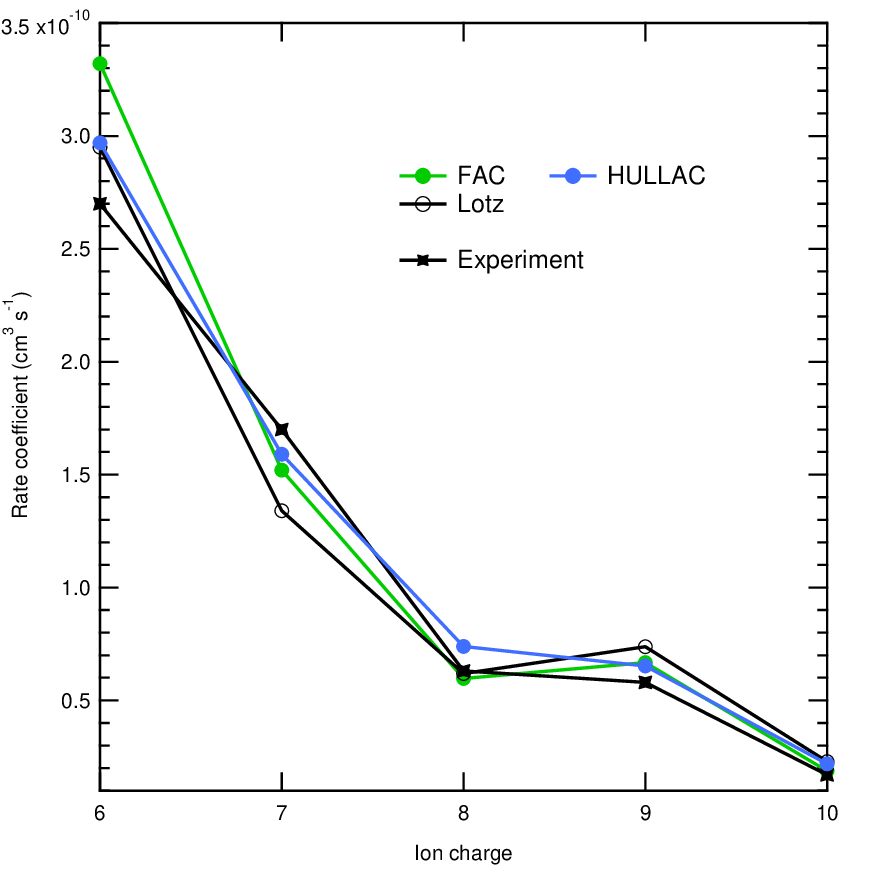}
\caption{EII rate coefficient of aluminum ions versus ion charge. Density and temperature: see Table \ref{Plasma conditions}. \label{Rate_comparison}}
\end{figure}

The fit procedure provides the $a_p$ coefficients and we are then able to calculate the rate coefficients within the Fermi-Dirac statistics (Eq. (\ref{qExpansion})) or Maxwell-Boltzmann approximation (Eq. (\ref{qprimeExpansion})). In our cases, a suitable fit is obtained with polynomials of order 5.

Figure \ref{Rate_comparison} represents the rate coefficients of different aluminum ions. Our calculations using HULLAC cross-sections are in good agreement with experimental results. This is also the case with the FAC code, except for N-like aluminum. The two calculations yield close rate coefficients. The difference between the rate deduced from the Lotz formula and the experimental value is  small. 

\section{Conclusion and prospective}
This work is devoted to the calculation of the ionization potential depression and the EII rates in plasmas. We focused our attention on aluminum plasmas at high density and astrophysical plasmas (CNO). Our calculations are compared to experimental results.

The Bingo code uses the robust classical molecular dynamics method. It allows to calculate the ionization potential depression accounting for all charge-charge interactions in the particle motion, within the limits of classical mechanics.
The choice of a regularized ion-electron potential which removes Coulomb divergence at short distances and accounts for some quantum effects, enables one to implement a ionization/recombination protocol to control the plasma ion charge distribution and the trapping of electrons in the ion wells.
Contrary to widely used methods (EK, SP, FLYCHK ...), it gives access to the ionization energy distribution function which considers the plasma perturbation and its fluctuating nature. Based on a statistical analysis of rare collisional events, the numerical determination of the ionization energy distribution function is very expensive and can be done mainly for the most probable populations of ion charges.
The simulation accounts for the ion dynamics but ignores the excited states and the radiative properties.

The IPD calculated within classical molecular dynamics (CMD) is compared to the models of Ecker-Kr\"oll and Stewart-Pyatt. At high density the CMD and Ecker-Kr\"oll IPDs are close. However, both calculations show a substantial discrepancy with experiment for the highest ion charge ($\simeq$ 30 eV). At low temperatures the CMD approach agrees with the formula of Ecker-Kr\"oll. When the temperature is increased the CMD IPD is closer to the formula of Stewart-Pyatt. It seems that the Ecker-Kr\"oll and Stewart-Pyatt formulas are two limits of the CMD model.

We have calculated the cross-sections and the rate coefficients in plasmas at near-solid density by using the Lotz formula where the IPD is taken into account. It is clear that the continuum lowering has an important effect on ionization by electron impacts.

In our plasma conditions (temperature and density) the free electrons degeneracy has a small effect on the ionization rate, which means that the Maxwell-Boltzmann approximation is satisfactory.

In a second work we investigated plasmas at lower densities. For such plasmas, the IPD is negligible.  Because the Lotz formula sometimes overestimates the cross-section, we introduce a new cross-section consisting in the product of the Lotz cross-section and a polynomial expansion whose variable is the ratio of the free electron energy to the ionization energy. The coefficients of the expansion are then adjusted in order to reach a good fit of the accurate cross-section given by two efficient atomic codes (FAC and HULLAC), in the DW approach. This new definition provides rate coefficients that are in better agreement with experimental values than is the Lotz formula. 

FAC and HULLAC are integrated software packages giving the atomic structure and cross-sections for collisional and radiative processes. Both are well-adapted to describe multicharged ion plasmas with configurations involving many open subshells giving rise to complex atomic structure. 
The main difference between FAC and HULLAC codes is that the first one uses a self-consistent potential and the second one a parametric potential.
Other methods (R-matrix and close coupling for instance) can provide accurate cross-sections. However they are not applicable to our cases due to large atomic level sets and to wide incident electron energy ranges.\\
\indent In the case of the CNO ions the measured cross-sections lie between the FAC and HULLAC results. The largest difference with experiment is of the order of 16 \% (for HULLAC, Oxygen case). The agreement between the experiment and the FAC code is better (less than 10 \%) than with HULLAC. The Lotz formula overestimates the CNO cross-sections.

More accurate cross-sections that are suitable for a comparison with experiments are in progress. We are considering a larger set of initial states from which electron ionization occurs. The agreement between our calculations and experimental results should be improved if we take into account the ionization from the metastable states. In fact, the populations of these states are of the same order of magnitude as the population of the ground level, as shown in the experiment of Fogle \textit{et al.} \cite{Fogle2008}. The contribution of the metastable states will be taken into account in a forthcoming publication.

Our calculated ionization rates of aluminum at low density ($\simeq 10^{16}$ electrons per cm$^3$) are compared to measurements. We used the new cross-section defined above but restricted ourselves to ionization from the ground state to all allowed excited states of the final ion. Our results show a better agreement with experiment than the Lotz formula. The agreement with experiment would be better if the ionization from excited states was also taken into account. In this case the population fractions of these levels are needed.

\appendix
\section*{Appendix A: Alternative method for the calculation of integral $\displaystyle\int_{\chi}^\infty \frac{\ln(\epsilon)}{e^{\epsilon-\eta}+1}d\epsilon$ \label{Appendix}}
Let $H(\epsilon)$ be any function varying smoothly with energy $\epsilon$ and $f(\epsilon)$ the Fermi-Dirac distribution 
\begin{equation}\label{FermiDirac}
f(\epsilon)=\frac{1}{e^{\epsilon-\eta}+1}.
\end{equation}
One has (see Ref.\cite{Grypeos1998}):
\begin{eqnarray}\label{Sommerfeld2}
\int_{\chi}^\infty H(\epsilon)f(\epsilon)d\epsilon=& &\int_{\chi}^\eta H(\epsilon)d\epsilon\nonumber\\
&+&\sum_{m=1}^\infty (-1)^{m}\left[\int_{\chi}^{\eta}H(\epsilon)e^{m(\epsilon-\eta)}d\epsilon-\int_{\eta}^{\infty}H(\epsilon)e^{-m(\epsilon-\eta)}d\epsilon\right],\nonumber
\end{eqnarray}
\textit{i.e.}
\begin{eqnarray}
\int_{\chi}^\infty \frac{\ln(\epsilon)}{e^{\epsilon-\eta}+1}d\epsilon=& &\int_{\chi}^\eta \ln\epsilon d\epsilon\nonumber\\
&+&\sum_{m=1}^\infty (-1)^{m}\left[\int_{\chi}^{\eta}\ln\epsilon~ e^{m(\epsilon-\eta)}d\epsilon-\int_{\eta}^{\infty}\ln\epsilon~ e^{-m(\epsilon-\eta)}d\epsilon\right],\nonumber
\end{eqnarray}
which gives the result

\begin{equation*}\int_{\chi}^\infty \frac{\ln(\epsilon)}{e^{\epsilon-\eta}+1}d\epsilon= \int_{\chi}^\eta \ln\epsilon d\epsilon
+\sum_{m=1}^\infty (-1)^{m}\left[A_m-B_m\right],\end{equation*}
with
\begin{eqnarray*}A_m&=&\int_{\chi}^{\eta}\ln\epsilon~ e^{m(\epsilon-\eta)}d\epsilon\\
&=&\frac{e^{-m\eta}}{m}\left[e^{\eta m}\ln\eta-e^{\chi m}\ln(\chi)+E_1(-\eta m)-E_1(-\chi m)\right]\end{eqnarray*}
and
\begin{equation*}B_m=\int_{\eta}^{\infty}\ln\epsilon~ e^{-m(\epsilon-\eta)}d\epsilon
=\frac{e^{\eta m}}{m}\left[E_1(\eta m)+e^{-\eta m}\ln\eta\right].\end{equation*}

\newpage

\providecommand{\newblock}{}

\end{document}